# Muzeel : A Dynamic JavaScript Analyzer for Dead Code Elimination in Today's Web

Tofunmi Kupoluyi
New York University Abu Dhabi
Abu Dhabi, UAE
jdk461@nyu.edu

Moumena Chaqfeh
New York University Abu Dhabi
Abu Dhabi, UAE
moumena@nyu.edu

Matteo Varvello
Nokia Bell Labs
varvello@gmail.com

Waleed Hashmi
New York University Abu Dhabi
Abu Dhabi, UAE
wah271@nyu.edu

Lakshmi Subramanian
New York University
NY, USA
lakshmi@cs.nyu.edu

Yasir Zaki
New York University Abu Dhabi
Abu Dhabi, UAE
yasir.zaki@nyu.edu

## ABSTRACT

JavaScript contributes to the increasing complexity of today's web. To support user interactivity and accelerate the development cycle, web developers heavily rely on large general-purpose third-party JavaScript libraries. This practice increases the size and the processing complexity of a web page by bringing additional functions that are not used by the page but unnecessarily downloaded and processed by the browser. In this paper, an analysis of around 40,000 web pages shows that 70% of JavaScript functions on the median page are unused, and the elimination of these functions would contribute to the reduction of the page size by 60%. Motivated by these findings, we propose *Muzeel* (which means *eliminator* in Arabic); a solution for eliminating JavaScript functions that are not used in a given web page (commonly referred to as *dead code*). *Muzeel* extracts all of the page event listeners upon page load, and emulates user interactions using a bot that triggers each of these events, in order to eliminate the dead code of functions that are not called by any of these events. Our evaluation results spanning several Android mobile phones and browsers show that *Muzeel* speeds up the page load by around 30% on low-end phones, and by 25% on high-end phones under 3G network. It also reduces the speed index (which is an important user experience metric) by 23% and 21% under the same network on low-end, and high-end phones, respectively. Additionally, *Muzeel* reduces the overall download size while maintaining the visual content and interactive functionality of the pages.

## 1 INTRODUCTION

The reuse of existing JavaScript code is a common web development practice which speeds up the creation and the amendment of web pages, but it requires sending large JavaScript files to web browsers, even when only part of the code is actually required. In this paper, we propose to eliminate JavaScript functions that are brought to a given web page, but never used. These functions are referred to as *dead code*.

The elimination of dead code is inspired by the fact that even though an unused function is never executed, it impacts the overall performance of the page because it must be processed by the browser. The impact of JavaScript is further worsened for users who solely rely on low-end smartphones to access the web [23]. For example, pages require overall triple processing time on mobile devices compared to desktops [12]. While methods like script-steaming (parsing in parallel to download) and lazy parsing can reduce JavaScript processing time, only a maximum of 10% improvement of the page load time is reported [19].

This work is motivated by an analysis of around 40,000 web pages, which shows that 70% of JavaScript functions on the median page are unused, and the elimination of these functions would contribute to the reduction of the page size by 60%. Given that user interactivity is a key feature that JavaScript provides in web pages, the dead code cannot be accurately identified unless all JavaScript functions that are executed when the user interacts with the page are reported.

However, the identification of dead code remains an open problem to date due to a number of challenges, including the dynamic nature of JavaScript that hinders the static analysis of the source code, the different ways in which the user can interact with a given page, and the highly dynamic changes in the structure and the state of modern web pages that occur due to user interactivity, which require handling the amended and the dynamically generated events accordingly.

The aforementioned challenges has led [13] to passively monitor JavaScript usage on web pages without capturing the functions that are triggered from the user interactivity, while the analysis in [24] ends once the page is loaded, implying that any JavaScript functions called afterward will not be considered. Instead of identifying only the dead code for potential elimination, [21] removes "less-useful" functions, a process that can drastically impact the functionality of interactive pages (more than 40% loss of page functionality can be witnessed when saving 50% of the memory).

To mitigate the impact of JavaScript on performance degradation of today's web pages [28], and to address the challenges of JavaScript dead code elimination in these pages, we design and implement *Muzeel*; a novel dead code analyzer that emulates the user interactivity events in web pages using a browser automation environment. *Muzeel* dynamically analyzes JavaScript code after the page load to accurately identify the *used* JavaScript functions that are called when the user interacts with the page (instead of a static JavaScript analysis that is proven to be inefficient [24]). It then eliminates the dead code of functions that are not being called by any of these events. *Muzeel* addresses the uncertainty of the user interactions (the various ways in which a user can interact with a given page), by covering potential combinations of the run-time interactions while considering the events dependency to trigger these events in the appropriate order. *Muzeel* can be applied to any web page, without imposing constraints at the coding style level.

We assume a medium CDN provider hosting the 40,000 most popular web pages [5]. We then dedicate a server machine to "crawl" these pages and run *Muzeel* to produce *Muzeel*-ed pages (with dead code eliminated). We finally setup a CDN *edge* node to serve both *original* and *Muzeel*-ed pages. The results show that for most pages the number of eliminated JavaScript functions ranges between 100 and 10,000, while the JavaScript size reduction ranges from 100KBytes to 5MBytes. Motivated by the aforementioned findings, we selected a set of 200 pages to evaluate the performance of *Muzeel* across different Android phones and browsers, and the quality of *Muzeel*-ed pages with respect of the original pages. Results show that *Muzeel* speeds up the page load by around 30% on low-end phones, and by 25% on high-end phones under 3G network. It also reduces the speed index (which is an important user experience metric) by 23% and 21% under the same network on low-end, and high-end phones, respectively. Additionally, *Muzeel* maintains the visual content and interactive functionality of most pages. The contribution of this paper are as follows:

- Analyzing JavaScript in 40,000 popular web pages, to investigate the number of unused functions and the potential of eliminating these functions on page size reduction. Our analysis shows that 70% of JavaScript functions on the median page are unused, and the elimination of these functions would contribute to the reduction of the page size by 60%.
- Proposing *Muzeel*; a novel solution for eliminating unused JavaScript functions in web pages through user interactivity automation that comprehensively considers the events dependency.
- Evaluating the impact of *Muzeel* on the page performance and quality using several Android mobile phones and browsers, showing significant speedups in page loads and speed index, and a reduction in page size by around 0.8 Megabyte, while maintaining the visual content and interactive functionality of most pages.

## 2 RELATED WORK
### 2.1 JavaScript Cost Mitigation

Web developers often utilize uglifiers (also known as minifiers) [9, 11] to reduce the size of JS files before using them in their pages. The reduction is achieved by removing unnecessary characters from JavaScript files, such as new lines, white spaces, and comments. Despite the enhancement in transmission efficiency (due to the reduced sizes of JavaScript files), the browser still has to interpret the entire JavaScript code, where a similar processing time is witnessed as in the case of the original code. In contrast, we mitigate the cost of JavaScript in web pages by completely eliminating unused code.

While methods like script-steaming (parsing in parallel to download) and lazy parsing can reduce JavaScript processing time, only a maximum of 10% improvement of the page load time is reported [19]. This is due to parsing one JavaScript resource at a time, while many other JavaScript resource are downloaded in parallel. With lazy parsing, some functions that are never executed are still being processed, for instance, when embedded in other functions that are about to execute.

In [24], the authors proposed to analyze JavaScript code to eliminate unused functions. This analysis was bounded by the time it takes a web page to load, which implies that any JavaScript functions called afterward, e.g., due to a user interaction or some dynamic JavaScript behavior, is ignored. The challenge we aim to address in this work is to identify *all* potential user events, independently of when the page loads, and then eliminate unused functions that are not being called when any of these events is triggered. More recently, the authors of [21] proposed to remove *less-useful* functions from JavaScript elements used in web pages to minimize memory usage in low-end mobile devices. They represent a given web page as a set of components structured as a dependency-tree graph and explored cutting in a bottom-up fashion, where more than 40% loss of page functionality can be witnessed when saving 50% of the memory.

## 2.2 Web Complexity Solutions

During the last decade, a number of solutions were proposed to speed up *complex* web pages. In [28], it is shown that JavaScript has a key impact on Page Load Time (PLT) due to its role in blocking the page rendering. To speedup PLT, Shandian [29] restructures the loading process of web pages, whereas Polaris [22] provides more accurate fetch schedules. However, the browser has to download and process all JavaScript elements brought by a given page. In a recent solution [10], a proxy server is implemented to offer a set of simplified web pages via a proxy server, where the identification of essential JavaScript elements was based on the JS access to web pages for reading, writing, or event handling. To prepare lightweight web pages, web developers can utilize Google Accelerated Mobile Pages (AMP) [15], which provides a set of restrictions in a web content creation framework. The impact of AMP pages on the user experience is characterized in [20]. While AMP offers the opportunity to create new lightweight pages, we aim to improve the browsing experience by optimizing JS usage in existing web pages.

## 3 MOTIVATION AND CHALLENGES

While methods like script-steaming (parsing in parallel to download) and lazy parsing can reduce JavaScript processing time, only a maximum of 10% improvement of the page load time is reported [19]. This is due to parsing one JavaScript resource at a time, while many other JavaScript resource are downloaded in parallel. With lazy parsing, some functions that are never executed are still being processed, for instance, when embedded in other functions that are about to execute.

The alternative approach proposed by *Muzeel* is to completely eliminate the code of the functions that are never executed (dead code), motivated by the high percentage of dead code in modern web pages. Our analysis of 25,000 web pages shows that 70% of functions in a median page are unused. The elimination of these functions would contribute to the reduction of the page size by 60%. A survey among 9,300 JavaScript developers rated dead code elimination as one of the highest requested features [16]. However, the identification and the elimination of JavaScript dead code from modern web pages is challenging and remains an open problem to date. In the remaining of this section, we briefly discuss these challenges.

## 3.1 The Dynamic Nature of JavaScript

The identification and the elimination of dead code is challenging and remains an open problem to date. This is basically due to the dynamic nature of JavaScript, which hinders the static analysis of the source code. More specifically, language features such as the possibility of dynamically accessing objects properties, and context binding that allows developers to assign an arbitrary object to "this" keyword, require a run-time analysis of the code to identify unused functions.

## 3.2 Handling User Interactivity

Given that user interactivity is a key feature that JavaScript provides in web pages, the dead code cannot be accurately identified unless all JavaScript functions that are executed when the user interacts with the page are reported. However, different run-time execution flows are expected due the uncertainty of the user interactions with the page. Consequently, a major challenge to address by a dynamic web-based JavaScript analyzer is to handle all combinations of interactive events in a given web page. This can be achieved by providing an accurate representation of the page elements along with their associated events, which requires the consideration of events dependencies across the page.

## 3.3 Handling Highly-Dynamic Pages

The dynamism of web pages adds another level of complexity to the aforementioned challenge. More specifically, additional interactive elements can be added dynamically to the page, and existing elements can be modified as the user interacts with the page. With this in mind, it is required to consider changes in the page structure and state to cover potential additional and/or modified events. This challenge has led [13] to analyze JavaScript usage from on web pages at page load, while the functions that are triggered when users interact with pages are not captured. Similarly, the analysis of dead code in [24] ends once the page is loaded, implying that any JavaScript functions called afterward will not be considered, and might be mistakenly eliminated.

## 3.4 Real-World Deployment Challenges

While a real world deployment would be helpful in identifying unused functions based on real user interactivity, it requires a representative number of users from different locations with different needs and interests, interacting with the page for sufficient periods of time. Instead of identifying only the dead code for potential elimination, [21] removes "less-useful" functions, a process that can drastically impact the functionality of pages (more than 40% loss of page functionality can be witnessed when saving 50% of the memory).

With *Muzeel*, we aim to accurately identify and eliminate unused JavaScript functions from web pages to avoid unnecessary browser processing. To handle the dynamic properties of JavaScript that hinder the identification of unused functions, we propose to dynamically analyze web pages by automating all possible user interactivity events in these pages. Since the user interactivity events are associated with JavaScript functions, *Muzeel* triggers each of these events and monitor the JavaScript functions that are called upon the occurrence of a given event. Any function that is never called by the page is considered as *unused function* and will be eliminated accordingly. We also propose a potential implementation that obtains and serves the optimized JavaScript files to achieve an improved user experience without impacting the pages' quality.

## 4 MUZEEL

*Muzeel* is a dynamic JavaScript analyzer which autonomously identifies dead code in web pages with the goal of improving the browsing experience, e.g., data savings and page speedups. *Muzeel* utilizes a black-box approach in which the unused JavaScript functions in web pages are identified without having knowledge of the JavaScript code and its implementation details. Unlike existing approaches, *Muzeel* runs dead code elimination autonomously without the need for "execution traces" [27] or real user interactions [14]. *Muzeel* addresses the aforementioned challenges by:

- Dynamically analyzing JavaScript code in web pages after the page load to accurately identify the dead code to eliminate, instead of a static JavaScript analysis that is proven to be inefficient [24].
- Thoroughly emulating users' interaction with a given web page, by automatically triggering all events present in the page, thus alleviating the need for a real user interactivity evaluation – a process that is time costly and prone to subjectivity.
- Addressing the uncertainty of the user interactions (the various ways in which a user can interact with a given page), by covering potential combinations of the run-time interactions while considering the events dependency (to trigger them in the appropriate order).

*Muzeel* is envisioned as a service offered by a CDN provider to help content owners optimize JavaScript code within their pages. JavaScript code used in today's web pages can be divided broadly into two categories: first party JavaScript, and third-party JavaScript. First party JavaScript refers to those files that are hosted within the same authoritative domain of the web page, e.g., by a CDN provider. Third-party JavaScript files refer to scripts hosted externally, outside the CDN; popular examples are Google Tag Manager/Analytics [3].

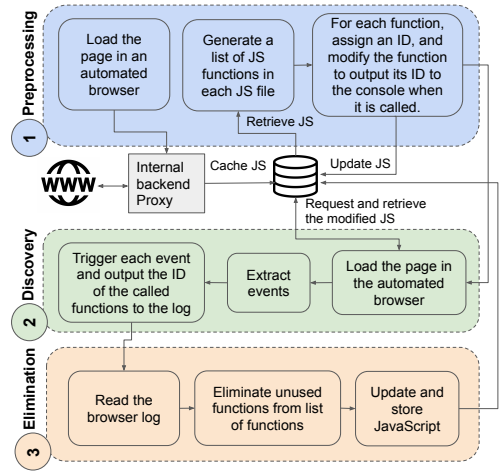

**Figure 1:** *Muzeel's* **architecture with the main processes.**

*Muzeel* eliminates dead code of first party JavaScript files, given the fact that the CDN is the authoritative entity responsible for hosting these files. For third-party JavaScript files, there are different scenarios offered by *Muzeel*, depending on the license of the JavaScript files, as well as the web page owner's preferences:

- For copyrighted JavaScript files, *Muzeel* does not perform the dead code elimination process, given that these files cannot be hosted by the CDN.
- For open-source/copylefted JavaScript files, *Muzeel* can perform the dead code elimination if the web page owner agrees to host local versions of these files.

Figure 1 shows *Muzeel's* architecture indicating the main processes: pre-processing, dead code discovery, and dead code elimination. Each of these processes is discussed further in this section.

### 4.1 Pre-Processing

The first phase of *Muzeel's* pre-processing focuses on creating an internal duplicate version of a given page on which the dead code elimination is carried out on. This version is used by *Muzeel* to modify the JavaScript files used by the page and dynamically analyze it in an automated browser environment. To create this duplicate version, *Muzeel* copies and hosts the web page along with its JavaScript files in an internal CDN back-end server.

In the second phase, a unique ID is assigned to each JavaScript function in each JavaScript file used by the page. The ID is represented in a form of the tuple <*file_name, function_startLine, function_endLine*>. A special console log call is added to the first line of every function which outputs its ID, that is, the file where the function is defined, as well as the function's start and end line numbers. *Muzeel* maintains a list of all IDs created during this process. These IDs cover all functions present in the JavaScript files used by the page.

When a page event triggers a particular JavaScript function, the function will print it's own ID to the browser's console log. Function IDs that are not printed to the browser's console log after triggering all page events are identified as *unused functions* or *dead code*.

### 4.2 Dead Code Discovery

Given the dynamic nature of JavaScript, discovering the used and unused functions by the page is not possible through static analysis [24]. As mentioned earlier (see Section 3), existing dynamic analysis methods do not consider the user interactions with the page such as clicking a button or navigating a drop-down menu. These interactions are crucial for identifying the JavaScript functions that are necessary for the page interactivity, given that these functions won't be under the dynamic analyzer radar unless their associated events are triggered.

The user interactivity on a given page is facilitated through events, such as hover, click, and focus. *Muzeel* identifies the functions required for user interactivity (*used functions*) through event listeners. An event listener monitors the occurrence of a certain event to call a JavaScript function or a chain of functions required to handle that event. Consequently, functions that are not called, either directly by any event listener or indirectly by other called-functions, are considered as *unused functions* or *dead code*. In order to trigger events, it is required to map these events to their corresponding page elements. A page element refers to an HTML tag that appears in the Document Object Model (DOM) of the page, such as image, button, or navigation element. *Muzeel* loads the page in a real browser environment and leverages the built-in functionality of that environment to identify the events and associate them to their corresponding page elements. Hence, *Muzeel* relies on the final DOM structure built by the browser without having to statically analyze the JavaScript code or create the DOM structure from scratch.

After identifying and mapping the events to page elements, *Muzeel* considers the events dependency to automatically trigger these events in the appropriate order through the browser automation environment (see Section 4.2.2). The functions called when an event is triggered are logged to the console. *Muzeel* monitors the console for log statements and obtains a list of functions that are called, in order to identify the used and unused functions. In the following, we discuss the design considerations in *Muzeel*'s dead code discovery.

*4.2.1 Page Elements Identification.* *Muzeel* uses the *XPath* to represent each page element and refer to it with a unique identifier. *XPath* is an accepted page element identification syntax on several browsers and browser automation tools. *Muzeel* chooses to use *XPath* for the following reasons:

- *XPath* allows for the extraction of the events associated with a specific element.
- *XPath* can also be used by browser automation tools to trigger events on a specific page element.
- *XPath* allows *Muzeel* to uniquely identify elements across reloads. This is crucial since *Muzeel* has to reload the page in some circumstances (for example, when redirected out of the page upon triggering a click event). *XPath* is used as opposed to using the internal representation of page elements provided by the browser automation tools, which may be invalid when the page is reloaded.

An *XPath* can be constructed using different strategies [1]: position-based, and attribute-based. The position-based strategy would fail in situations such as adding/removing an element to/from the page. Additionally, the attribute-based strategy requires the presence of unique attributes which identify every element, where such attributes might not always be available. Consequently, *Muzeel* considers both approaches in constructing the *XPath*, where in the presence of a tag element id, *Muzeel* uses the attribute-based strategy given that these ids are unique. Whereas, in the absence of the tag element id, *Muzeel* reverts back to the position-based approach, where the *XPath* is constructed using position-based indexing from the nearest parent element with an "id" or "class" attribute.

In an HTML document representing a given web page, elements are structured internally using different tags. Each tag can have a unique "id", and a reference to a pre-defined "class" of attributes from the accompanying Cascading Styling Sheets (CSS) files, which set the different visual attributes for a given tag. To demonstrate this, given a page:

```
1  <html>
2      <body>
3          <div id="div1">
4              <button></button>
5          </div>
6          <div class="divClass2">
7              <a style="color:blue;"></a>
8          </div>
9      </body>
10 </html>
```

**Listing 1: A sample page to demonstrate XPath construction in Muzeel**

The "<button>" will be constructed using the hybrid strategy from the "<div>" element with id, "div1" and is identified as: $//div[@id = "div1"]/button[1]$. This "<div>" element, since it possesses an "id" attribute, is simply represented as:

//𝑑𝑖𝑣[@𝑖𝑑 = "𝑑𝑖𝑣1"], using just the attribute-based strategy. The second "<div>" element, since it possesses a "class" attribute, is also represented as: //𝑑𝑖𝑣[@𝑐𝑙𝑎𝑠𝑠 = "𝑑𝑖𝑣𝐶𝑙𝑎𝑠𝑠2"]. Lastly, although the "<a>" element contains a "style" attribute, *Muzeel* does not consider "style" to be an identifying attribute. Consequently, the "<a>" element is represented using the hybrid strategy from the "<div>" with class, "divClass2", //𝑑𝑖𝑣[@𝑐𝑙𝑎𝑠𝑠 = "𝑑𝑖𝑣𝐶𝑙𝑎𝑠𝑠2"]/𝑎[1].

4.2.2 *Emulating User Interactivity.* To emulate user interactivity on a given web page, *Muzeel* utilizes a browser automation environment to trigger all interactive events on the page, with the consideration of all commonly used page interactive events, including but not limited to: *mousedown, mouseup, mouseover, mouseout, keydown, keypress, keyup, dblclick, drag, dragstart,* and *dragend*.

To obtain a comprehensive list of these events on the page, *Muzeel* uses the *XPaths* generated for all page elements. From the *XPath*, all events attached to a given element are extracted and added to a list of events. Due to events dependency, some events can only be successfully triggered after the occurrence of another set of events. For example, the elements under the "More" drop-down menu shown in Figure 2 are only interactive after clicking "More". In addition, triggering a certain event may prevent the interactivity with another event, such as an event that causes a modal to open on a page where the modal blocks the elements appearing behind it and prevents the interactivity with these elements (see Figure 3). Therefore, *Muzeel* considers both the event dependency as well as the page state changes in emulating the user interactivity.

Algorithm 1 summarizes the user interactivity emulation in *Muzeel*. First, the list of events is obtained to store the interactivity events in *eventList* (line 4). Each "Event" is an object with the following attributes: the *eventType* (which refers to the type of event), the *XPath* of the element the event is associated with, the parent event *parentEvent* that is required to be triggered before the event can occur, and a list of dependent or successor events *successorEvents*. When *Muzeel* gets the event from the *XPath*, the *parentEvent* is yet to be determined.

To determine events dependency, *Muzeel* leverages a breadth-first search to identify the parent event for every event in the *eventList*. It instantiates an *eventQueue* with a *baseEvent*, which is an ancestor of all events. When the *baseEvent* is triggered, no action is performed. While the *eventQueue* is not empty (line 12), *Muzeel* pops the first event from the *eventQueue* and stores it in the *parentEvent*. *Muzeel* seeks to find all elements in *eventList* whose parent is *parentEvent*. *Muzeel* does this by triggering the *parentEvent*, and determining which other events can be successfully triggered after the *parentEvent*. Once an event is successfully triggered, it is removed from the *eventList* and added to the *eventQueue*.

---

**Algorithm 1** Muzeel User Interactivity Emulation

1: **INPUT** : *xpathList*[]
2: **OUTPUT** : *baseEvent*
3: **procedure** TRIGGEREVENTS(*xpathList*)
4:    **List<Event>** *eventList*;
5:    **for** *xpath* **in** *xpathList* **do**
6:        *xpathEvents* = getEventsFromXpath(*xpath*);
7:        **for** event **in** *xpathEvents* **do**
8:            *eventList*.add(*event*);
9:    **Event** *baseEvent*;
10:   **Queue<Event>** *eventQueue*;
11:   *eventQueue*.add(baseEvent);
12:   **while** *eventQueue* ≠ ∅ **do**
13:       *parentEvent* = *eventQueue*.pop();
14:       refreshPage()
15:       triggerEvent(*parentEvent*);
16:              ▷ predecessorEvents are triggered first
17:       **for** *event* **in** *eventList* **do**
18:           **if** triggerEvent(*event*) == True **then**
19:               *parentEvent*.addChild(*event*)
20:                      ▷ predecessorEvents are assigned
21:               *eventQueue*.add(*event*)
22:               *eventList*.remove(*event*)
23:               refreshPage()
24:               triggerEvent(*parentEvent*)
25:   **return** *baseEvent*

---

The first *parentEvent* to consider is the *baseEvent*, which results in no changes in the page state when triggered. The *successorEvents* of *baseEvent* are those events that can be directly triggered on the page load. It is important to note that before triggering a given event, *Muzeel* triggers all its predecessor events. To determine *successorEvents* of *baseEvent*, *Muzeel* loops through all the events in the *eventList* to find events that can be triggered from the current page state. Any event that can be triggered will be added as an element in the *successorEvents* list of *baseEvent*. In doing so, the *baseEvent* is assigned as the *parentEvent* of the added event. The event is then removed from the *eventList* so it is not triggered again by *baseEvent*. It is then added to *eventQueue* so that in a subsequent iteration of the while loop (line 12), we can find all events which have this added event as their *parentEvent*. After this is done, we refresh the page and trigger the *baseEvent* to return to the page state of *baseEvent* (a fresh page load). This is required to overcome the aforementioned challenge of events blocking other events and to ensure that an accurate event dependency is determined. A full event-dependency graph is drawn from the *baseEvent* returned by this process.

*4.2.3 Addressed Challenges.* Web pages may contain "hidden elements" that are not visible or interactive on the first page load and would only appear upon the occurrence of a certain event. For example, the navigation elements of a drop-down menu may not be visible until the menu button is clicked or hovered over. *Muzeel* addresses the issue of hidden elements in web pages, by accurately considering the events dependency such that the necessary predecessor events are triggered in the right order to reveal the hidden element and trigger any successor event afterwards (Section 4.2.2).

A common functionality in web pages is the open/close mechanism that characterizes some elements such as interactive menus. These elements can open up hidden components when clicked, and close these components when clicked again. Therefore, opening and closing can change the page state, such that the interactivity state of the page elements when a given component is opened differs from their state when that component is closed. To handle this issue, *Muzeel* triggers all click events three times, where the 1st click captures the function(s) behind opening the hidden component, the 2nd click captures the function(s) behind closing the hidden component, and the 3rd click reopens the hidden component so that the user interactivity with the successor elements included in the component can be emulated. If the same function handles opening/closing, then it will be logged twice, however, *Muzeel* discards duplicate log statements.

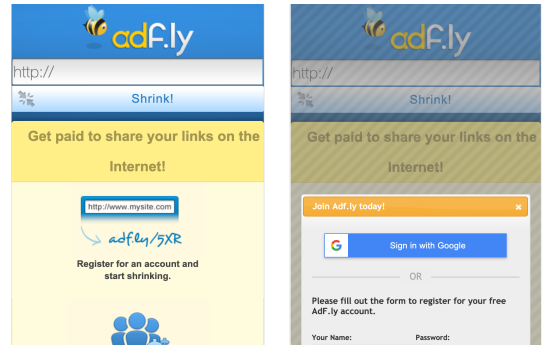

(a) The original state of adf.ly when no events are triggered

(b) adf.ly after a button click which opens a modal that blocks the elements behind

Figure 3: An example showing a different page state in (b) in comparison to the original state shown in (a) after a click event is triggered.

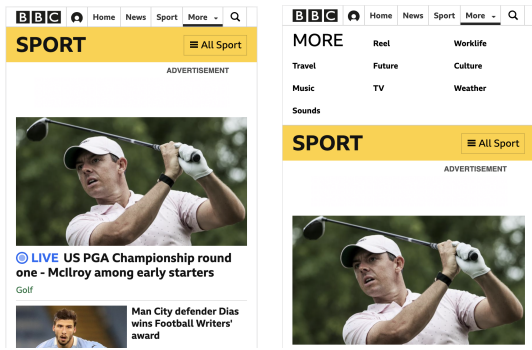

(a) bbc.com/sport with a hidden element rooted at "More" menu button

(b) bbc.com/sport when "More" menu button is clicked

Figure 2: An example showing a hidden page element

## 4.3 Dead Code Elimination

When the dead code discovery process completes, *Muzeel* uses the browser's console logs – which contains the IDs of the JavaScript functions that are called – to annotate the *used* functions. Consequently, the functions that are never called can be determined and removed from their respective JavaScript files. The simplified JavaScript files (with the unused functions eliminated) are saved into the database to be served with the page instead of the original JavaScript files.

It is worth mentioning that *Muzeel* implicitly considers nested functions. Specifically, when triggering a given event leads to calling a nested function, the latter function is also called, and both are reported to the browser's console. Similarly, in a case where a function is removed, all its subsequently nested functions are also removed. This means that even though *Muzeel* does not preserve the hierarchy amongst the functions in the initial constructed list, it successfully obtains a complete trace of the called functions, and therefore, the functions that are not called can be accurately identified as *unused functions*.

## 5 IMPLEMENTATION
## 5.1 CDN Back-end Proxy

Given that an actual CDN deployment is challenging, we implement the next most realistic scenario for the purpose of evaluation. We assume a medium-large CDN provider hosting the first 150,000 web pages [5] in Alexa's top 1M list. We then dedicate a powerful server machine (equipped with 64 cores and 1 TB of RAM) to crawl a set of web pages (while recording the full HTTP(S) content and headers). For each page, two copies of JavaScript files are stored – an original version, and a version to be optimized by *Muzeel*, which we refer to as the *Muzeel version* of the JavaScript file. This version of the JavaScript file will be served by the proxy for the purpose of dead code elimination. It is updated at two points, firstly during pre-processing (when the function log calls are added), and secondly, during dead code elimination (when the unused functions are removed). Caching and serving pages are achieved by extending the mitmproxy [8] which intercepts regular browser traffic and serves it locally, emulating the role of a CDN edge node.

## 5.2 Browser Automation

*Muzeel* uses Selenium [18] to automate the browser environment in order to perform the dead code elimination. Selenium is a framework for automated web testing, where developers can use a set of internal APIs to load web pages through different browsers, simulate and control the user-interactions with the page, access the different page elements and events, and modify the page DOM structure. *Muzeel* uses the Selenium Chrome web-driver, where Chrome is chosen due to its popularity as well as its accompanying Chrome DevTools Protocol (CDP) that is vital in obtaining page events and mapping them to their corresponding individual elements. *Muzeel* loads a given web page in the automated browser, and then extracts a full list of page elements from the DOM structure generated by the browser along with the event listeners [2] associated with the page elements using the CDP. For every element, *Muzeel* generates an *XPath* identifier to locate each element as described in Section 4.2.1, identify elements across reloads, and facilitate the emulation of the user interactivity with these elements.

## 5.3 Identifying Page Elements and Extracting Events

When *Muzeel* loads a given web page via Selenium, it extracts the HTML DOM structure of the page rendered by the Chrome browser. It then uses *jsoup* – a popular open source HTML parser, in the process of elements identification (see Section 4.2.1) to parse the page DOM and construct the *XPaths* of the page elements (see Section 4.2.1) from a Depth First Search (DFS) traversal.

*Muzeel* uses the CDP to extract the events associated with a given *XPath*. This is achieved by first obtaining the "objectId" associated with the *XPath* – the objectId is Chrome's internal id for elements on the page. *Muzeel* obtains the objectId using the Chrome Dev Tools, "document.evaluate" call. *Muzeel* then passes the obtained objectId to Chrome Dev Tools "DomDebugger.getEventListener" functionality to extract the eventListener associated with the given objectId. This process is carried out for all the elements on the page. A list of events associated with an element is returned from this call. *Muzeel* takes each event returned and creates an Event object for each with the *XPath* and eventType.

## 5.4 Emulating User Interactivity via an Interaction Bot

To emulate the user events, *Muzeel* implements an *Interaction Bot* that leverages Selenium bundled with Chrome Web-driver. *Muzeel* navigates through the events retrieved in 5.3, taking the events dependency into account (as described in Section 4.2.2). For each event, *Muzeel* locates the associated element using the *XPath*, and then attempts to trigger the event on that element. It relies on the native implementation to trigger the event in case the event behavior is implemented in Selenium (such as the case of *click*, *dblclick*, and *drag*). On the other hand, in cases where a specific event is not available in Selenium, the event is simulated using a collection of available Selenium user-interaction events. For example, to emulate a user moving in and out of the element as specified by a *mouseout* event, Selenium's *moveToElement(element)* function is first called (which moves the mouse cursor to the element's position), then, *moveByOffset(x,y)* is called (which moves the cursor some arbitrary offset away from where the element is located).

Events that are not related to user interactions are not explicitly triggered — such as the "load" event that is triggered when page load is completed. This is because these events are usually triggered implicitly by the browser, and therefore do not need to be explicitly triggered for their associated functions to be logged.

# 6 MUZEEL EVALUATION

*Muzeel*'s evaluation revolves around the potential of the dead code elimination over a large dataset, performance (page timing metrics, page size, number of requests), resource utilization (CPU and battery savings), and interplay with different browsers, phone types, and networking conditions.

## 6.1 Methodology

Given that an actual CDN deployment is challenging, we deploy *Muzeel* using the next most realistic scenario. We assume a medium CDN provider hosting the 40,000 most popular web pages [5] from Alexa's top 1M list. We then dedicate a powerful server machine (equipped with 64 cores and 1 TB of RAM) to "crawl" the landing pages of these 40,000 web pages. The pages are loaded via Chrome while recording the full HTTP(S) content and headers using mitmproxy [8]. Next, we produce *Muzeel*-ed pages by running our dead code discovery and elimination mechanism. Finally, we setup a CDN *edge* node (10 ms latency to the user when assuming a fast WiFi) which can serve both *original* and *Muzeel-ed* pages. This is achieved using mitmproxy which intercepts regular browser traffic and serves it locally, emulating the role of a CDN edge node. Testing devices are regular Android phones where mitmproxy's root CA (Certificate Authority)[1] is installed to properly handle HTTPS. Note that this step is only required for testing purposes as an actual CDN owns the certificates for the domains it serves.

| Name | Vendor | OS | CPU Info | Memory | Battery |
|---|---|---|---|---|---|
| Redmi Go | Xiaomi | Android 8.1 Oreo (Go edition) | Quad-core 1.4 GHz Cortex-A53 | 1GB RAM | Li-Ion 3000mAh |
| SMJ337A | Samsung | Android 8.0.0 | Quad-core 1.4 GHz Cortex-A53 | 2GB LPDDR3 | 2,600mAh |
| SM-G973F/DS | Samsung | Android 9.0 (Pie) | | 8GB RAM | Li-Ion 3400mAh |

Table 1: *Muzeel* test-bed composition

*6.1.1 Datasets.* To evaluate the potential of *Muzeel* (see Section 6.2), we use the full set of 40,000 cloned web pages. On the other hand, to evaluate *Muzeel*'s performance (see Section 6.3), we selected 200 web pages from the 40,000 that we have previously cloned and *Muzeel*-ed. These pages were selected as follows. First, we consider the 1,500 most popular pages from the full data-set. Next, we divided the 1,500 pages into four buckets by exponentially increasing the bucket size (*i.e.,* doubling the bucket size every step), staring with a bucket size of 100 and ending with a bucket size of 800. Then, from each bucket we uniformly chose 50 pages.

*6.1.2 Client-side Test-bed.* The client-side test-bed consists of several Android mobile devices – from low end (Xiamoi Redmi Go, Samsung J3) to high end (Samsung S10) – whose characteristics are summarized in Table 1. For most of our tests we rely on Chrome, as it is today's most popular browser. We also experiment with Edge and Brave (which was chosen for its raising popularity with a current 25 million monthly users [7], and advanced ad-blocking capabilities [4]). The phones connect to the Internet over a fast WiFi – with a symmetric upload and download bandwidth of about 100 Mbps; when needed, network throttling was used to emulate different cellular networks.

*6.1.3 Network Settings.* We emulate three cellular networks:

- 3G: this represents a slow cellular network with a download bandwidth of 1.6Mbps, upload bandwidth of 768 Kbps, and a round trip time of 300ms.
- LTE: this represents a moderate cellular network with a download/upload bandwidths of 12Mbps, and a round trip time of 70ms.
- LTE+: this represents a fast cellular network with a download bandwidth of 42Mbps, upload bandwidth of 25Mbps, and a round trip time of 40ms.

*6.1.4 Browser Automation and Performance Metrics.* Each mobile device connects via USB to a Linux machine which uses the `WebPageTest` browser automation tool. This tool is used to automate both web page loads and telemetry collection, e.g., performance metrics and network requests. We focus on classic web performance metrics [25] (FirstContentfulPaint, SpeedIndex and PageLoadTime), as well as CPU, and bandwidth usage. For the Samsung J3, we also report on battery consumption measured by a power meter directly connected to the device in battery bypass [26]. Given that not all browsers on Android allow communication with their developer tools, which is used by `WebPageTest`, we have also developed a tool which uses the Android Debugging Bridge (adb) [6] to automate a browser, *i.e.,* launch and load a webpage, while monitoring resource utilization. We then leverage `visualmetrics`[2] to extract performance timing metrics from a video of the web page loading.

## 6.2 The Potential of Muzeel

We start by studying the ability of *Muzeel* to identify and eliminate JavaScript dead code. We consider the following metrics: the numbers of eliminated JavaScript functions, JavaScript size reduction, time required per page, and "running frequency", which measures how frequently *Muzeel* should run depending on how quickly web pages change.

We used *Muzeel* to perform the dead code elimination on the top 40,000 webpages from Alexa's top 1M list. On our powerful server, this process took about 2 days, assuming 10 threads, or about 6 core per web page. Figure 4(a),4(b) show the per-JavaScript file distribution of the eliminated number of JavaScript functions and their corresponding size in bytes, respectively. The outer figures show the histograms computed for a total of 100 bins, whereas the inner figures show the Cumulative Distribution Functions (CDFs). It can be seen from Figure 4(a) that 20% of JavaScript files (the upper $20^{th}$ percentile of the CDF) had more than 100 eliminated functions, with cases reaching above 10,000 eliminated functions. The CDF also shows that, on median, the number of eliminated JavaScript functions per file is about 12.

Additionally, Figure 4(b) shows that around 70% of JavaScript files have an eliminated size between 1 - 100 KBytes (with a small percentage of going beyond 10MBytes). It's worth mentioning here, that most of these files are generally small, and for a lot of them *Muzeel* only removes the functions' bodies but keeps the functions' headers intact. On the other hand, we computed the percentage of eliminated size in comparison to the original JavaScript file size, and found that in about 10% of the JavaScript files *Muzeel* eliminated about 98-100% of the file size. The CDF shows that the size of the eliminated JavaScript files is about 5 KBytes at the median. To understand *Muzeel*'s dead code elimination on the overall page rather than the individual JavaScript files, we computed both the number of eliminated functions and the eliminated bytes on a per page bases. The results are shown in Figures 4(d) 4(e), respectively. Figure 4(d) shows a Normal distribution for the per-page number of eliminated JavaScript functions, with a mean around 20,000 eliminated functions. The results show that for most pages the number of eliminated JavaScript functions ranges between 100 and 10,000. This is a significant deduction in the number of unused JavaScript functions which can be eliminated without impacting the pages content or functionality (see Section 6.3.3). Figure 4(e) shows the size reduction in bytes of the above JavaScript dead code elimination on a per page basis. The results show a similar Normal distribution with a mean of about 0.8MBytes. It can also be seen that for most of the pages the JavaScript size reduction ranges from 100KBytes to 5MBytes, which further strengthen the potential of *Muzeel*.

---

[1]https://docs.mitmproxy.org/stable/concepts-certificates/

[2]https://github.com/WPO-Foundation/visualmetrics

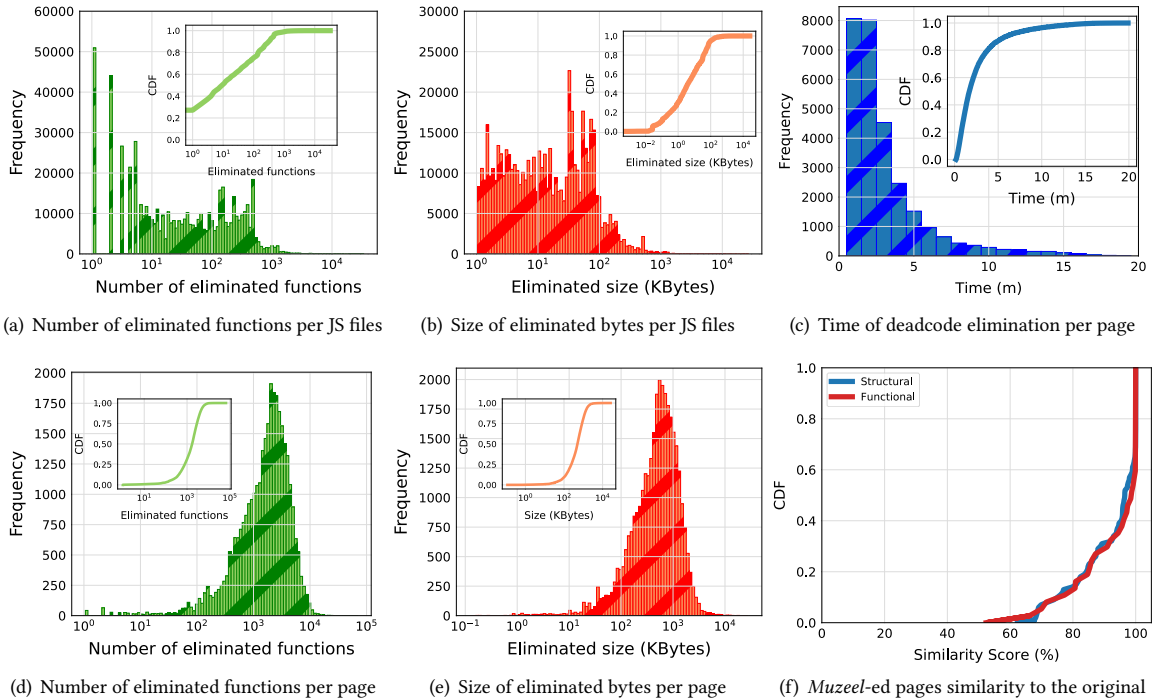

Figure 4: Dead code elimination statistics computed across 40,000 popular web pages[3]

[3]Apart from (f) which is computed across 200 web pages.

To assess the time complexity of *Muzeel*, we compute the time taken to perform the dead code elimination for each of the web pages. Figure 4(c) shows the histogram (and CDF) of the aforementioned time in minutes. The figure shows that for about 85% of the pages, *Muzeel* requires at most 5 minutes. Note that this duration was obtained assuming up to 6 cores used concurrently. For the remaining 15% of the pages, we measured a duration of up to 20 minutes. This result suggests that *Muzeel* can easily run each time a web page is updated, to ensure the correctness of the JavaScript dead code elimination. On our powerful server, by dedicating all 64 cores to the process, this process would take about 30 seconds, on median.

Finally, we assess the quality of the pages with the dead code elimination in terms of both the structural similarity and the functional similarity with respect to the original versions. Figure 4(f) shows the CDF of the scores computed using *PQual* [17], which is a tool that compares web pages to compute the structural and the functional scores using computer vision. The structural similarity CDF, depicted in blue, shows that for 70% of the pages *Muzeel* maintains a similarity score of above 90%. For the lowest scoring 10% pages *Muzeel* has a similarity score between 70-80%. Similar observation can be seen in the functional similarity score of

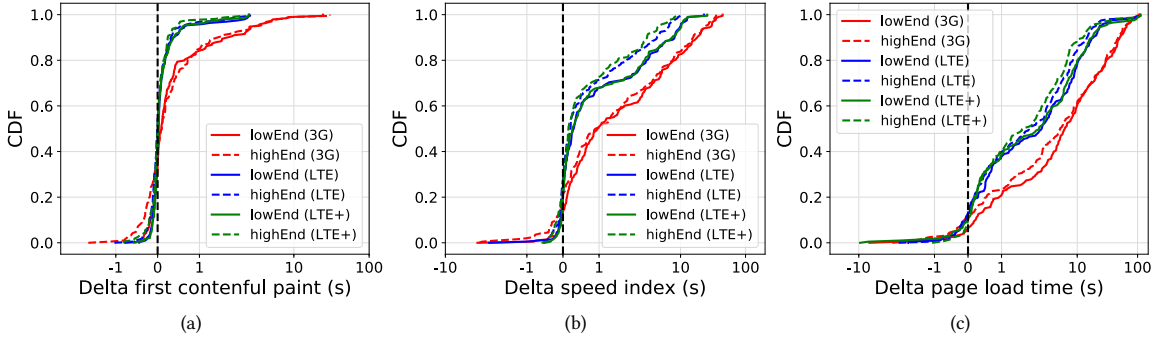

Figure 5: Delta performance results using different networks and phone types

| Network | Phone | PLT | | | SpeedIndex | | | Dom complete | | |
|---|---|---|---|---|---|---|---|---|---|---|
| | | % | Muzeel | Original | % | Muzeel | Original | % | Muzeel | Original |
| 3G | LowEnd | 31.5 | 31.7 | 46.3 | 23.4 | 8.5 | 11.1 | 22.7 | 23 | 29.8 |
| 3G | HighEnd | 25.8 | 33.7 | 45.4 | 21.5 | 8.6 | 10.9 | 21.6 | 23.1 | 29.5 |
| LTE | Low-End | 27.3 | 11.2 | 15.5 | 19.5 | 3.8 | 4.8 | 17.9 | 7.9 | 9.6 |
| LTE | High-End | 30.1 | 9.4 | 13.5 | 16.3 | 3.3 | 4 | 16.7 | 6.6 | 7.9 |
| LTE+ | Low-End | 29.2 | 9.7 | 13.7 | 22.1 | 3.4 | 4.3 | 20.9 | 6.2 | 7.8 |
| LTE+ | High-End | 25.6 | 7.8 | 10.5 | 12.2 | 3 | 3.4 | 17.7 | 4.8 | 5.9 |

Table 2: *Muzeel*'s median results

*Muzeel*, depicted in red, where the CDF almost matches the structural similarity CDF.

## 6.3 Muzeel Performance

In contrast to the previous evaluation that focused on highlighting the percentages of eliminated JavaScript dead code, here, we study the impact of the eliminated dead code on the overall user experience. For this evaluation, we use three Android mobile devices (see Table 1) and the 200 web pages described in Section 6.1.1. We compare the performance of these 200 pages with the dead code eliminated by *Muzeel* with respect to their original versions. Each page was loaded 5 times for each version, and for each metric we consider the median out of the 5 runs.

### 6.3.1 Network-Based Evaluation.
Figure 5 shows the CDFs of the delta performance results of the *Muzeel*-ed pages with respect to their original versions, in terms of the aforementioned timings metrics, using both the low-end and the high-end phones, under three emulated networks: 3G, LTE, and LTE+. For each page and metric, the delta is computed by subtracting the value of that metric measured for the *Muzeel*-ed page from the value measured for the corresponding original page. It follows that values bigger than 0 represents *Muzeel* savings, while values smaller than zero represent penalties. Experiments were conducted using Chrome.

FirstContentFulPaint (FCP) is a web quality metric capturing the first *impression* of a website, which many users often associate with what defines a web page "fast" [25]. Figure 8(a) shows three trends, regardless of the network condition and the device: some web pages (20-30%) show minimal slow down (average of few hundred ms), some exhibit no performance difference (up to 40% on LTE and LTE+), and the majority (up to 60% in 3G) show significant FCP speedups, up to several seconds. Intuitively, *Muzeel* speedups arise from JavaScript which are saved before FCP. Given this metric is quite fast, and JavaScript tend to be loaded later in a page, it is expected to see many web page with equivalent FCP between their original and *Muzeel*-ed version. More unexpected are the few negative results. Their explanation lies in the intricacy of the web – and the HTTP protocol itself – where removing or shrinking some objects can change the order of requests, e.g., by anticipating a larger object even if not contributing to a specif metric. The figure also shows much higher speedups and slowdowns when considering 3G; this is expected given that the lower bandwidth inflates the differences between the two loading strategies.

Next, we focus on SpeedIndex (SI) a web quality metric which aims at capturing the "average" user experience [25]. Compared with FCP, Figure 5(b) shows a clear shift to the right, with 70-80% of the pages benefiting from some speedups. This happens because SI is an overall "later" metric which gives more chances to *Muzeel* to offer its savings. The same trend is also confirmed in Figure 5(c), which instead focuses on the PageLoadTime (PLT), or the time at which a browser fires the onLoad event, suggesting that all content has been loaded. In this case, *Muzeel* offers speedups for 90-95% of the web pages. With respect to the networking conditions, both figures confirm the previous trend with much higher deltas in the presence of 3G. With respect to the mobile devices, the figures show higher benefit for the low-end, likely due to a reduction in CPU usage as we will discuss later. The median values of *Muzeel*'s performance are shown in Table 2.

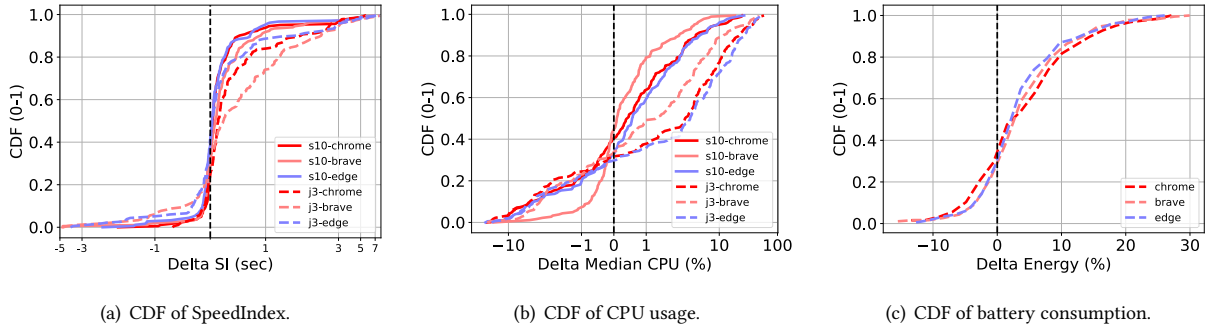

(a) CDF of SpeedIndex.

(b) CDF of CPU usage.

(c) CDF of battery consumption.

Figure 6: Delta performance results using different browsers and phone types

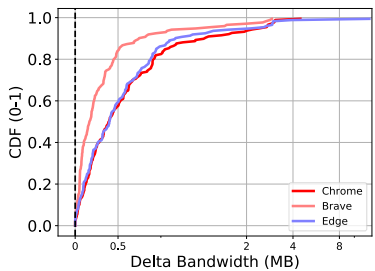

Figure 7: CDF of bandwidth consumption across browsers: Brave, Chrome, Edge.

6.3.2 *Browser-Based Evaluation.* In the previous section, *Muzeel*'s web performance was assessed across a variety of network conditions. Here, we focus on various browsers: Chrome, Edge, and Brave. In this evaluation, we study the resource utilization (CPU, bandwidth, and battery when possible) and also report on SpeedIndex (SI). Experiments were run both on a high-end and a low-end device; differently from before, we replace the low-end device with a Samsung J3 which we have previously connected to a power meter for fine-grained battery monitoring.

Figure 7 shows the CDF of the delta bandwidth utilization (original - *Muzeel*-ed) across browsers. The figure shows, overall, very similar savings across devices for the same browser, which is expected and thus confirm correctness in the experimentation technique. The figure confirms that Chrome and Edge are very similar browsers, and indeed *Muzeel* achieves equivalent bandwidth savings on both browsers: median of about 400KB, up to multiple MBytes. When considering Brave, the bandwidth savings from *Muzeel* are reduced by about 50%. This happens due to the lack of tracking and advertisement code – mostly JavaScript – which Brave removes via its integrated adblocker, thus giving less a chance to *Muzeel* to provide savings. However, even in the case of Brave *Muzeel* realizes data savings in the order of MB for 10% of the web pages. Note that Brave runs a very aggressive adblocker [4], and thus these numbers represent a lower bound on the expected data savings provided by *Muzeel* in presence of adblocking.

Following up from the previous result, we next investigate how fast (or slow) *Muzeel* would make the user experience across browsers and devices. We only report on SpeedIndex (SI) given that it is the web timing metric which captures the "average" end-user experience and for which the previous section has shown "average" performance improvements. Accordingly, Figure 6(a) shows the CDF of SpeedIndex per browser and device. The figure shows an overall trend similar to the previous result (see Figure 5(b)), with about 70% of the websites showing performance improvements of up to several seconds. Next, 10-20% of the pages (according to browser and device) shows minimal slow down of a maximum of 150ms, followed by a longer tail which can reach up to 5 seconds. The figure shows that Brave benefits from most performance improvements, which is counter-intuitive given the previous result on the bandwidth savings. We conjecture that this additional improvements originate from the reduced load on Brave's adblocker, a complex task whose extra cost is significant especially on low-end devices, which is usually amortized by bandwidth savings. In this case, *Muzeel* helps Brave's adblocker by achieving similar bandwidth savings with less computation cost from the device.

Next, we evaluate *Muzeel*'s impact on CPU consumption. We sample CPU usage once per second during a web page load, and then report the median consumption. Accordingly, Figure 6(b) shows the CDF of the (delta) median CPU utilization across browsers and devices. As above, Chrome and Edge – on a given device – achieve very similar trends. Differently from before, the figure shows a larger fraction of websites (up to 45% in the case of Brave on J3) for which *Muzeel* causes extra CPU usage, up to a 10% increase. However, note the 20-30% of these websites are within a 1% CPU increase which is just too small to be statistically significant (same holds for an even larger fraction of a 1% CPU decrease, as purposely highlighted by the x-axis). Larger CPU degradation is instead associated with websites with massive speedups (more than one second for 10-20% of the websites) where *Muzeel* compresses the overall page load in a shorter time causing a temporary burst in CPU usage. Last but not least, the figure shows overall less CPU variation, either positive or negative, in the case of Brave, regardless of the device. This happens for two reasons: 1) Brave is a lean browser with overall smaller CPU consumption than, for instance, Chrome, 2) less impact due to the lacks of ads, as shown by Figure 7.

Finally, we focus on the J3 device only and comment on battery consumption. We use a power meter to derive the mAh consumed during a original and *Muzeel*-ed web page load, and then compute their delta. Given that mAh are hard to be related to the actual savings/penalty, we then report the result as a percentage of the battery consumption of the original version. For about 70% of the pages we observe energy savings ranging from 0 to 30%, regardless of the browser.

### 6.3.3 Comparison To State-of-the-art.
Here, we compare *Muzeel* performance and pages quality to the dynamic analyzer of *Lacuna* [24], where the results are shown in Figure 8. In summary, *Muzeel* outperforms Lacuna in terms of FPC, SI, PLT, network requests reduction, page size reduction, as well as the similarity to the original pages as shown in Figures 8(a), 8(b), 8(c), and 8(d), respectively. Specifically, in comparison to Lacuna, *Muzeel* improves the median PLT by more than 7 seconds and the SI by 0.5 seconds. Figure 8(a) shows that in about 80% of the pages, *Muzeel* improves the timing metrics over Lacuna, without sacrificing page content and functionality as observed in Figure 8(d). Additionally, *Muzeel* has a page size savings across all evaluated pages in comparison to Lacuna, with a median of 350KBytes and a maximum of 10MBytes size reductions. Figure 8(d) shows the qualitative evaluation comparison between *Muzeel* and Lacuna, where it can be seen that *Muzeel* structural (solid blue curve) and functional (dashed blue curve) outperforms Lacuna (depicted in red). *Muzeel* maintains a structural and functional score close to 100% for more than 50% of the pages, in comparison to a score of around 82% in Lacuna.

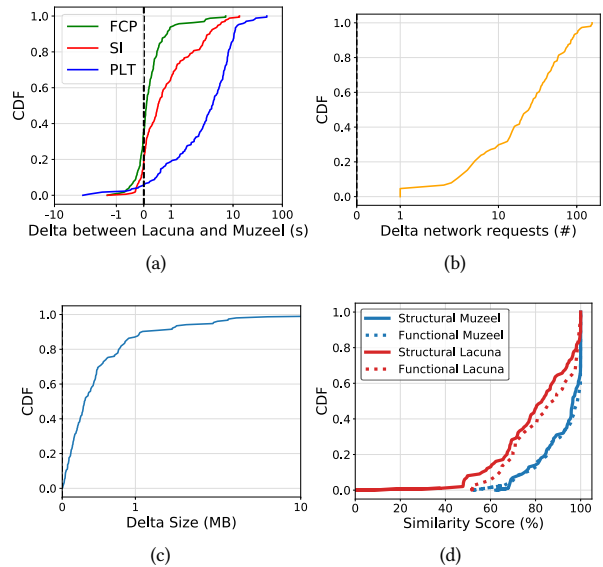

Figure 8: Performance comparisons to Lacuna

### 6.3.4 Dead Code Elimination Frequency.
We evaluated how frequently should we run *Muzeel* to eliminate the dead code, by cloning a set of 60 web pages spanning different categories (news, education, sports, business, commercial, entertainment) every 12 hours in a period of a week. Results show that 99% of the JavaScript after dead code elimination does not change over a period of a week. This means that CDF providers do not need to re-run *Muzeel* with frequency less than one week.

## 7 CONCLUSION
In this paper, we analyzed 40,000 web pages to quantify unused JavaScript in today's web. Motivated by our findings, we proposed and evaluated *Muzeel*, which provides significant speed ups in the page loads and reductions in page sizes across different Android phones and browsers.

## REFERENCES

[1] [n.d.]. https://www.w3.org/TR/xpath/
[2] [n.d.]. Chrome DevTools Protocol. https://chromedevtools.github.io/devtools-protocol/tot/DOMDebugger/#method-getEventListeners

[3] 2009. Google Analytics. https://analytics.google.com/analytics/web/. Accessed: 2021-04-23.

[4] 2020. Which is the Best Free Ad Blocker? https://brave.com/learn/best-ad-blocker/. Accessed: 2021-05-26.

[5] 2021. 2020 state of the CDN industry. https://blog.intricately.com/2020-state-of-the-cdn-industry-trends-market-share-customer-size. Accessed: 2021-03-19.

[6] 2021. Android Debug Bridge (adb). https://developer.android.com/studio/command-line/adb. Accessed: 2021-05-26.

[7] 2021. Brave Passes 25 Million Monthly Active Users. https://brave.com/25m-mau/. Accessed: 2021-05-26.

[8] 2021. A Free and Open Source Interactive HTTPS Proxy. https://mitmproxy.org/. Accessed: 2021-04-28.

[9] Mihai Bazon. 2012. UglifyJS. http://lisperator.net/uglifyjs/. Accessed: 2020-05-01.

[10] Moumena Chaqfeh, Yasir Zaki, Jacinta Hu, and Lakshmi Subramanian. 2020. JSCleaner: De-Cluttering Mobile Webpages Through JavaScript Cleanup. In *Proceedings of The Web Conference 2020*. 763–773.

[11] CircleCell. 2011. JSCompress - The JavaScript Compression Tool. https://jscompress.com/. Accessed: 2020-05-01.

[12] Houssein Djirdeh. 2019. JavaScript | 2019 | The Web Almanac by HTTP Archive. https://almanac.httparchive.org/en/2019/javascript. Accessed: 2020-01-2.

[13] Utkarsh Goel and Moritz Steiner. 2020. System to Identify and Elide Superfluous JavaScript Code for Faster Webpage Loads. *arXiv preprint arXiv:2003.07396* (2020).

[14] Utkarsh Goel and Moritz Steiner. 2020. System to Identify and Elide Superfluous JavaScript Code for Faster Webpage Loads. *arXiv preprint arXiv:2003.07396* (2020).

[15] Google. 2019. AMP is a web component framework to easily create user-first web experiences - amp.dev. https://amp.dev. Accessed: 2019-05-05.

[16] Sacha Greif. 2016. The state of JavaScript survey. https://stateofjs.com/. Accessed: 2021-05-5.

[17] Waleed Hashmi, Moumena Chaqfeh, Lakshmi Subramanian, and Yasir Zaki. 2020. PQual: Automating Web Pages Qualitative Evaluation. In *The Adjunct Publication of the 33rd Annual ACM Symposium on User Interface Software and Technology* (Virtual (previously Minneapolis, Minnesota, USA)) *(UIST '20)*. Association for Computing Machinery, New York, NY, USA.

[18] Jason Huggins. 2019. Selenium WebDriver. Browser Automation. https://www.seleniumhq.org/projects/webdriver/. Accessed: 2019-05-14.

[19] Marja Hölttä and Daniel Vogelheim. 2015. New JavaScript techniques for rapid page loads. https://blog.chromium.org/2015/03/new-javascript-techniques-for-rapid.html. Accessed: 2021-05-4.

[20] Byungjin Jun, Fabián E Bustamante, Sung Yoon Whang, and Zachary S Bischof. 2019. AMP up your Mobile Web Experience: Characterizing the Impact of Google's Accelerated Mobile Project. In *The 25th Annual International Conference on Mobile Computing and Networking*. 1–14.

[21] Usama Naseer, Theophilus A Benson, and Ravi Netravali. 2021. WebMedic: Disentangling the Memory-Functionality Tension for the Next Billion Mobile Web Users. In *Proceedings of the 22nd International Workshop on Mobile Computing Systems and Applications*. 71–77.

[22] Ravi Netravali, Ameesh Goyal, James Mickens, and Hari Balakrishnan. 2016. Polaris: Faster Page Loads Using Fine-grained Dependency Tracking. In *13th USENIX Symposium on Networked Systems Design and Implementation (NSDI 16)*. USENIX Association, Santa Clara, CA. https://www.usenix.org/conference/nsdi16/technical-sessions/presentation/netravali

[23] Ravi Netravali and James Mickens. 2018. Prophecy: Accelerating Mobile Page Loads Using Final-state Write Logs. In *15th USENIX Symposium on Networked Systems Design and Implementation (NSDI 18)*. USENIX Association, Renton, WA, 249–266. https://www.usenix.org/conference/nsdi18/presentation/netravali-prophecy

[24] Niels Groot Obbink, Ivano Malavolta, Gian Luca Scoccia, and Patricia Lago. 2018. An extensible approach for taming the challenges of JavaScript dead code elimination. In *2018 IEEE 25th International Conference on Software Analysis, Evolution and Reengineering (SANER)*. IEEE, 291–401.

[25] Karolina Szczur. 2020. Performance | 2020 | The Web Almanac by HTTP Archive. https://almanac.httparchive.org/en/2020/performance. Accessed: 2021-05-26.

[26] Matteo Varvello, Kleomenis Katevas, Wei Hang, Mihai Plesa, Hamed Haddadi, Fabián E Bustamante, and Benjamin Livshits. 2019. BatteryLab, a distributed power monitoring platform for mobile devices: demo abstract. In *Proceedings of the 17th Conference on Embedded Networked Sensor Systems*. 386–387.

[27] Hernán Ceferino Vázquez, Alexandre Bergel, S Vidal, JA Díaz Pace, and Claudia Marcos. 2019. Slimming javascript applications: An approach for removing unused functions from javascript libraries. *Information and Software Technology* 107 (2019), 18–29.

[28] Xiao Sophia Wang, Aruna Balasubramanian, Arvind Krishnamurthy, and David Wetherall. 2013. Demystifying Page Load Performance with WProf. In *Presented as part of the 10th USENIX Symposium on Networked Systems Design and Implementation (NSDI 13)*. USENIX, Lombard, IL, 473–485. https://www.usenix.org/conference/nsdi13/technical-sessions/presentation/wang_xiao

[29] Xiao Sophia Wang, Arvind Krishnamurthy, and David Wetherall. 2016. Speeding up Web Page Loads with Shandian. In *13th USENIX Symposium on Networked Systems Design and Implementation (NSDI 16)*. USENIX Association, Santa Clara, CA, 109–122. https://www.usenix.org/conference/nsdi16/technical-sessions/presentation/wang